\documentclass[manuscript,screen,table,acmsmall]{acmart}
\AtBeginDocument{%
  }

\usepackage{CJKutf8}
\usepackage[table]{xcolor}
\usepackage{multirow}
\usepackage{appendix}
\usepackage{longtable}

\newcommand{\wzy}[1]{{\color{black} #1}}
\newcommand{\cjl}[1]{{\color{black} #1}}

\newcommand{\zhiyang}[1]{{\color{black} #1}}

\newcommand{\zyw}[1]{{\color{black} #1}}
\setcopyright{acmlicensed}
\copyrightyear{2026}
\acmYear{2026}
\acmDOI{}
\begin{document}

\title[Understanding Legal Information Seeking on Video-Sharing Platforms]{``Law at Your Fingertips'': Understanding Legal Information Seeking on Video-Sharing Platforms in China}

\author{Zhiyang Wu}
\email{zhiyangwu5-c@my.cityu.edu.hk}
\affiliation{%
  \institution{City University of Hong Kong}
  \city{Hong Kong}
  \country{China}
}

\author{Junliang Chen}
\affiliation{%
  \institution{Xi'an Jiaotong University}
  \city{Xi'an}
  \country{China}}
\email{}

\author{Qian Wan}
\affiliation{%
  \institution{City University of Hong Kong}
  \city{Hong Kong}
  \country{China}
}

\author{Qing Xiao}
\affiliation{%
 \institution{Carnegie Mellon University}
 \city{Pittsburgh}
 \state{Pennsylvania}
 \country{United States}}

\author{Piaohong Wang}
\affiliation{%
  \institution{City University of Hong Kong}
  \city{Hong Kong}
  \country{China}}

\author{Ge Gao}
\affiliation{%
  \institution{University of Maryland}
  \city{College Park}
  \state{Maryland}
  \country{United States}}

\author{Zhicong Lu}
\affiliation{%
  \institution{George Mason University}
  \city{Fairfax}
  \state{Virginia}
  \country{United States}}
\email{zlu6@gmu.edu}


\begin{abstract}

\zyw{Equipping laypeople with the capabilities to seek legal information has been an important goal for \textit{Legal Empowerment} in modern society. However, unlike general information-seeking behaviors, legal information seeking is characterized by high stakes, urgency, and a critical need for emotional support, which traditional text-based searching platforms struggle to satisfy. In recent years, people have been increasingly turning to Video-Sharing Platforms (VSPs) for access to legal information and to fulfill their legal needs. Despite the importance of this shift, such VSP-mediated legal information-seeking practices remain underexplored. Through an observational analysis of legal content on two VSPs (Douyin and Bilibili) and interviews with 20 Chinese information seekers, this study examined the practices and challenges associated with seeking, comprehending, and evaluating legal information on VSPs. We further revealed the formation of trust and engagement on the VSP-based legal knowledge-sharing community, highlighting how VSP affordances helped mitigate seekers' epistemic discomfort and satisfy their needs for emotional support. In the discussion, we provided insights on balancing \textit{heuristic and systematic processing} to encourage information cross-validation, and offered implications for designing trustworthy civic information systems and fostering an accessible, safe, and efficient information-seeking environment in digital space.}

\end{abstract}





\maketitle

\section{INTRODUCTION}

\zyw{

\textit{Legal Empowerment} has been an important goal pursued in modern society with the call for equitable access to justice for all society members~\cite{golub2010legal, domingo2014politics, van2008access, sepulveda2015beyond}. Due to the specialized knowledge requirements for the legal domain, including legal textual literacy and a nuanced understanding of various legal principles, it has been particularly challenging for laypeople to properly seek and comprehend legal information~\cite{borg2013legal, van2008access}. Therefore, a major mission in \textit{Legal Empowerment} is to better equip laypeople with the capabilities to seek legal information and utilize the knowledge to realize their rights \cite{PLEDefinition, IDLODefinition}.

However, unlike general lifestyle information seeking online, which has been extensively documented by prior HCI research (e.g., food recipe~\cite{wang2025s}, health advice~\cite{dewan2024teen, augustaitis2021online}, video games~\cite{hirsh2012reality}), legal information seeking presents fundamentally different characteristics and challenges. Firstly, in the legal domain, decisions made by stakeholders often carry significant risks, and misinterpreting legal information can result in severe and irreversible consequences, including financial ruin, loss of liberty, or the forfeiture of rights~\cite{Xiong2023duan, song2023lawfluencers, crowe2023real}. Moreover, past research has demonstrated that legal information seeking is rarely proactive, but often driven by distressful events (e.g., divorce, labor dispute, debt) and approaching court deadlines~\cite {pleasence2014people}. Therefore, legal information-seeking behaviors are often characterized by high anxiety and urgency~\cite {pleasence2014people}, where information seekers expect higher epistemic authority of the information sources, more comprehensible and actionable advice, and more empathetic emotional support compared with seeking information in other domains~\cite{vischer2006legal, ipsos2017analysis, sandefur2020legal}. However, conventional online platforms, which have historically centered on text-based Q\&A forums (e.g., Quora, Reddit) or static government repositories (e.g., China Judgments Online), often fail to address these needs of legal information seekers due to their high requirements of legal textual literacy and ability for jurisdiction verification in law~\cite{thomas2014online, deady2000cyberadvice, franklin2001hanging}.

In the digital era, the development of social media, particularly video-sharing platforms (VSPs), has shown unique potential to empower laypeople and address the challenges associated with seeking legal information. 
The accessibility and synchronicity of VSPs offer a more convenient, affordable, and efficient approach to receiving legal advice~\cite{Xiong2023duan}. With turning to VSPs becoming the first choice of an increasing number of legal information seekers~\cite{Xiong2023duan}, we have been witnessing the reshaping of how citizens access justice in the digital age. However, despite the importance of this shift, existing HCI and CSCW literature lack a comprehensive understanding of the practices and challenges faced by legal information seekers on VSPs. By studying how they seek, comprehend, and evaluate legal information, and how they interact with legal content creators and peers on VSPs, we can not only enhance our understanding of the algorithmic mediation of legal expertise, but also inform the design of trustworthy civic information systems for better supporting equitable access to justice in the digital space. Therefore, we aim to address the following research questions (RQs) in this study:

\begin{itemize}

    \item {\textbf{RQ1}: How do legal information seekers utilize VSP affordances to fulfill their specific legal needs?}

    \item {\textbf{RQ2}: How do interactions with legal content creators and peers shape the formation of trust on VSPs?}

    \item {\textbf{RQ3}: What strategies do legal information seekers employ to evaluate the quality of legal information on VSPs?}
             
\end{itemize}

To answer these questions, we observed and characterized the legal content on two popular VSPs in China (Douyin and Bilibili), followed by in-depth semi-structured interviews with 20 Chinese legal information seekers. We uncovered 
that VSPs empower Chinese legal information seekers with their unique affordances, including emerging live legal consultations that provide synchronous and interactive counsel, as well as various types of legal videos that promote legal knowledge to laypeople. Aligning with previous studies on social media influencers~\cite{chou2023like}, we found that viewers' initial impression of legal content creators helps establish their parasocial trust in them. More importantly, however, what uniquely shaped seekers' engagement with VSP-mediated legal community is the relaxing, friendly, and empathetic communication atmosphere, which significantly mitigated their anxiety and epistemic discomfort towards the legal system, and satisfied their needs for emotional support. Despite the accessibility and engagement, however, we argued that the affordances of VSPs prioritize \textit{heuristic processing} over \textit{systematic processing}~\cite{chaiken2012theory}, which can make viewers less sensitive to the quality of information. 
With a relatively low willingness to verify legal information on VSPs and limited cross-validation approaches, we discussed the challenges for laypeople's evaluation practice and highlighted the need for VSP designs to encourage deep, systematic processing and critical thinking.
We concluded our study with design implications for supporting a more trustworthy, efficient, and empathetic legal information-seeking experience on VSPs.

This study contributes to HCI and CSCW mainly by: (i) comprehensively revealing current practices of VSP-mediated legal information seeking, and reflecting on the potential and concerns of VSP affordances in supporting legal empowerment; (ii) uncovering the formation and engagement of the VSP-based legal knowledge-sharing community, highlighting the influence and significance of VSP mediation of expertise; (iii) providing insights into the challenges of legal information evaluation on VSPs, and offering implications for deigning trustworthy civic information systems and fostering an accessible, safe, and efficient information seeking environment in digital space.
}

\section{RELATED WORK}

\zyw{In Section~\ref{literature-information-seeking}, we presented past research efforts on information-seeking behaviors on various social media platforms, explained the unique characteristics and challenges of legal information seeking, and articulated the significance of studying legal information seeking on Video-Sharing Platforms (VSPs). In Section~\ref{literature-legal-empowerment}, we first explained the concept of \textit{Legal Empowerment} and its relationship to \textit{Legal Information Seeking}. We then summarized the literature on the development of accessible legal information in the past few decades.}

\subsection{Information Seeking on Social Media Platforms}
\label{literature-information-seeking}

Information seeking on social media is a growing area of HCI research, examining how users utilize these platforms to access, filter, and verify information. While social media platforms such as Twitter, Facebook, and Instagram have become important channels for Western users to obtain information due to their immediacy, interactivity, and personalized recommendation systems~\cite{cao2021you,humphreys2017social, dewan2024teen,milton2024seeking}, for Chinese people, they are more familiar with domestic substitutions like Douyin (Chinese version of Tiktok), Xiaohongshu (sometimes referred to as ``Chinese Instagram''), Sina Weibo (microblogging platform), and Bilibili (a comprehensive video-sharing site)~\cite{fitzgerald2022chinese,lu2019fifteen}. \zyw{With the popularity of these social media platforms, they have become the primary channels for people around the world to seek advice on issues related to their daily lives or self-development, including healthcare~\cite{augustaitis2021online,dewan2024teen,milton2024seeking,starling2016information, zhang2022shifting}, career development~\cite{blaising2019navigating, tomprou2019career, malik2024towards}, food recipe~\cite{wang2025s}, video games~\cite{hirsh2012reality}, and other general lifestyle areas~\cite{jonas2024we,nicholson2019if}. However, compared to other domains, legal information seeking presents fundamentally different characteristics and challenges, yet remains largely unexplored in the HCI literature. In the legal domain, decisions made by stakeholders often carry significantly more risks than seeking information for entertainment or daily-life advice~\cite{Xiong2023duan, song2023lawfluencers}. Legal information is often about life-altering problems (e.g., divorce, crime, debt), and misinterpreting it can result in severe and irreversible consequences, including financial ruin, loss of liberty, or the forfeiture of rights~\cite{Xiong2023duan, song2023lawfluencers, crowe2023real}. Moreover, unlike proactive exploratory information-seeking behaviors (e.g., seeking a cook recipe for a good dinner), past research has demonstrated that legal information seeking is often driven by distressful legal problems and approaching court deadlines~\cite{pleasence2014people}. Therefore, legal information-seeking behaviors are often characterized by extreme anxiety and urgency~\cite {pleasence2014people}, where information seekers have a higher requirement for epistemic authority of the information sources and a critical need for more empathetic emotional support~\cite{vischer2006legal, ipsos2017analysis, sandefur2020legal}. }

In recent years, video-sharing platforms (VSPs) such as YouTube, Bilibili, and TikTok (Douyin) have become unprecedentedly popular and attracted billions of active users around the world~\cite{bartolome2023literature, hughes2024viblio}. With numerous legal video creators and live streamers sharing their content on these platforms, people increasingly consider VSPs as influential sources of legal information~\cite {liu2024modeling, niu2023building}. \zyw{The unique affordances of VSPs, such as synchronicity and performativity, hold the potential to facilitate laypeople's access to legal information by offering a more efficient and interactive approach to receiving legal advice~\cite{Xiong2023duan}, but are also accompanied by risks of misinformation and quality evaluation. While people expect high accuracy and credibility of information on social media~\cite{chou2023like}, the perceived quality can significantly influence their judgment on the utility of information~\cite{knight2005developing}. Therefore, when laypeople are dealing with information that requires professional analysis to verify its actual quality (e.g., legal information), they often fail to evaluate it effectively due to a lack of specialized knowledge~\cite{dewan2024teen, zhang2022shifting}. For example, Dewan et al.~\cite{dewan2024teen} reported that when seeking information about reproductive legal requirements, teens in the U.S. found it hard to distinguish misinformation on social media. 
Prior studies have shown that content on VSPs suffers from inconsistent quality and can potentially spread misinformation or controversial content due to the low barriers to publishing videos or opening live-streaming channels on such platforms~\cite{chou2023like, hughes2024viblio, niu2023building}. Moreover, the "participatory culture" on many VSPs allows non-experts to share their opinions on legal issues without much verification~\cite{hughes2024viblio, niu2023building}, which poses even greater risks to the quality of legal information on VSPs.

Given that legal information seeking presents different characteristics and challenges compared with general lifestyle information-seeking behaviors, and the unique potential and risks brought by the VSP affordances, our study aims to develop a deeper understanding of how legal information seekers access, comprehend, and evaluate
legal information on VSPs, as an attempt to provide insights into VSP mediation of expertise and designing more trustworthy civic information systems.
}



\subsection{Legal Empowerment and Accessible Legal Information}
\label{literature-legal-empowerment}


With the increasing call for justice and human rights in the modern world, \textit{Legal Empowerment} has been an important goal pursued by governments and many non-governmental organizations~\cite{golub2010legal, domingo2014politics, van2008access, sepulveda2015beyond}. 
Legal empowerment can be defined as ``Strengthening the capacity of all people to exercise their rights, either as individuals or as members of a community'' \cite{PLEDefinition}, which emphasizes the importance of accessible legal information and legal advice in modern society~\cite{PLEDefinition, IDLODefinition}. A key mission in legal empowerment is to equip people with the capabilities to access and utilize legal information, thereby enabling them to realize their rights \cite{PLEDefinition, IDLODefinition}. 
Just as the International Development Law Organization argues, ``\textit{rights mean little if those entitled to them are unaware they exist}'' \cite{IDLODefinition}.
However, many people live outside the law's protection, especially in developing countries, who generally have lower incomes~\cite{borg2013legal, van2008access}. \zyw{To better protect human rights and enhance the legal literacy of the general public, modern society requires more accessible and affordable methods for laypeople to seek legal information.}

\zyw{In the 20th century of China, many citizens faced great difficulties accessing and comprehending legal information~\cite{michelson2008dear}. In the 1980s, people in China typically acquired legal information by reading legal books and some legal advice columns in newspapers \cite{givens2019justice}. An example would be the legal advice column \textit{Dear Lawyer Bao}, which published over 460 legal advice articles with cases between 1989 and 1998 in the \textit{Beijing Evening News} \cite{michelson2008dear}. 
Entering the 21st century, with the development of information and communication technologies (ICTs), access to legal information has increasingly shifted to online websites and text-based forums~\cite{thomas2014online}. In Western contexts, researchers have put considerable effort into understanding the legal advice consulting behaviors on text-based websites (e.g., Quora, Reddit)~\cite{lanctot2002regulating, turnbull2018communicating, maggs2006free, givens2019justice, bickel2015online, thomas2014online, denvir2016online}, which are platforms where clients post questions for legal professionals to answer. Similarly, on one of China's largest legal consultation websites, \textit{findlaw.cn}, more than 9.5 million questions were posted from 2012 to 2017, among which more than half of the questions were accompanied by some answers \cite{givens2019justice}. Moreover, people can refer to official government repositories, such as \textit{China Judgments Online}, for real judgment cases. Compared to making appointments with attorneys in their offices, the convenience and cost-effectiveness of posting questions or searching for information online quickly garnered the attention of both lay people and legal experts \cite{thomas2014online, deady2000cyberadvice, franklin2001hanging}. 
However, despite great improvement compared to traditional lawyer-client interactions,  these text-based online forums or static websites still struggle to meet all the needs of legal information seekers. For example, seeking information on these platforms requires legal
textual literacy and the ability for asynchronous verification in law~\cite{thomas2014online, deady2000cyberadvice, franklin2001hanging}, which challenges laypeople without legal expertise. Concerns were also raised about the accountability of such asynchronous interactions, as what seemed merely suggestions to legal experts may be perceived as legal advice to laypeople, and many answers lacked essential details~\cite{thomas2014online}. The lack of face-to-face interactions also made it difficult for legal information providers to establish trust with information seekers and to provide them with proper emotional support~\cite{szczepanska2020seeking}, which is a critical need in legal information seeking~\cite{cheong2024not}.

In recent years, the development of video-sharing platforms (VSPs) has shown unique potential to empower laypeople and address these challenges associated with conventional legal information-seeking platforms. The emerging legal livestreaming consultations and various legal videos offer information seekers more engaging and synchronous interactions with legal content creators and peers. In this study, we aimed to conduct the first investigation into how these VSP affordances uniquely shape legal information-seeking practices and critically analyze the implications of this new phenomenon towards the broader goal of legal empowerment and equitable access to justice.

}

\section{METHODOLOGY}

To answer the research questions, we first observed legal content on several popular VSPs in China, followed by conducting semi-structured interviews with 20 participants who had experience seeking legal information on VSPs. The observation provided us with more contextual information on the typical formats of legal content that information seekers may encounter on these platforms, which inspired the design of the question for subsequent interviews with information seekers. This study protocol was approved by our Institutional Review Board (IRB).

\subsection{Observing Legal Content on VSPs}

Through the observation process, we aimed to gain insights into i) the typical formats of legal content on VSPs and ii) the affordances of VSPs that support viewers' legal information seeking.

\subsubsection{Selection of Platforms}
\label{plat-selection}
To conduct this study, we examined various popular VSPs in China that potentially support legal information seeking, including Douyin, Bilibili~\cite{he2023seeking}, Kuaishou~\cite{lu2019feel}, and Xiaohongshu (also known as RedNote). We found that, compared to other platforms, Douyin and Bilibili have the largest number of legal video creators and legal live streamers, and their legal content comprehensively covers various legal fields (e.g., crime, employment, marriage). Moreover, these two platforms have the top legal content creators across the Chinese Internet, such as \textit{``Luo Xiang Talks about Criminal Law''} (\begin{CJK*}{UTF8}{gbsn}罗翔说刑法\end{CJK*}, Bilibili, 30 million followers) and \textit{Lawyer Long Fei} (\begin{CJK*}{UTF8}{gbsn}龙飞律师\end{CJK*}, Douyin, 9 million followers). Therefore, we selected Douyin and Bilibili as the representative platforms for conducting our observation.

\subsubsection{Observing Legal Live Streaming Sessions and Short Videos}

To comprehensively understand the existing legal content shared on the chosen video-sharing platforms, we conducted two rounds of observation of legal live-streaming sessions and short videos. Before observing, the three researchers met to discuss the details of conducting the observation, and they agreed on the following schemes:

\textbf{For live streaming sessions}: We searched with keywords ``legal live streaming'', ``legal content creators'', or ``legal consultations'' on the selected video-sharing platforms. \zhiyang{We went to the ``livestreaming'' section of the search results, which contained all the currently streaming sessions, and checked the sessions one by one.} We chose the live-streaming sessions which indicated the live streamers were offering legal advice to the audiences, for example, directly stating ``Legal Consultations on Crime'' or ``Specialized in Resolving Loan Disputes'' in the streaming session titles. \zhiyang{Livestreaming sessions with more audiences were prioritized in observation for increased opportunities of catching ongoing consultations. Moreover, to cover a wider range of legal topics in the observation, we might exclude some sessions that discuss similar topics to the previously observed ones. We also excluded streaming sessions that primarily promoted their offline services rather than offering online consultations.} For each selected live stream, we watched the whole process of one or two consultations for each session (typically 20 - 30 minutes). When observing, the researchers monitored and took notes of the
interactions between live streamers and information-seekers, including their verbal and textual communications and the platform affordances they used, as an attempt to understand how information-seekers conveyed their legal needs to the live streamers and how the VSPs supported such live-stream consultations. We also paid close attention to the legal questions raised by the information seekers and the corresponding legal fields (e.g., Civil and Commercial Laws, Criminal Laws). At the end of each observation, we took a screenshot of the live-streaming session we observed as a record (see~\autoref{fig:ls-screenshot} for an example).

\textbf{For legal videos}: \zhiyang{We aimed to collect videos that offered legal information, such as legal knowledge, legal advice, or legal news comments, etc.} Therefore, we searched with keywords ``legal videos'', ``legal knowledge'', or ``legal news comments'' on the selected video-sharing platforms. \zhiyang{On Douyin, we set the ranking algorithm as ``comprehensive ranking'', which was the default ranking mode on Douyin, so that we may mitigate the effect of biased recommendations based on personal preference and cover a wider range of videos accessible to ordinary information seekers. On Bilibili, we chose ``default ranking'' algorithm for similar reasons. We did not impose restrictions on the video length, number of views, or publish date to reach a wider range of legal videos}. We selected the videos that typically represent the legal opinions of creators and can potentially be considered information sources by viewers, including those that offer legal information, provide legal advice, or comment on current legal news. These videos often indicated their relevance to legal information in the title, such as ``Teach You how to Deal with Contract Disputes'' and ``Do You Know these Behaviors are Illegal?'', and they typically lasted around 2-3 minutes. \zhiyang{Similar to the practice in live streaming sampling, in order to cover a wider range of legal topics in the observation, we might exclude some videos that were discussing similar topics as the previously observed ones. We also excluded videos that promoted offline legal services or commercial products, such as legal books, rather than sharing legal knowledge or information.} For each video observed, we summarized and took notes about the general information of the videos, including their presentation forms and the legal topics discussed. To further understand the behavior of information seekers while watching these videos, we observed the comments made by viewers under the video content. This content may indicate whether information seekers perceive the videos as providing useful information or knowledge. Viewers may also raise their questions in the comments, hoping to receive answers from the creators or other viewers. We also observed some debates on controversial legal cases in the comments. These information seekers' behaviors helped us understand the viewer's community and the ecosystem for legal information-seeking on VSPs. At the end of each observation, we recorded the link to the video.

\zhiyang{We further categorized the observed legal topics in live-streaming sessions and short videos based on the Chinese law system~\cite{chineselaws}, and provided example legal questions discussed during the live sessions or within the short videos (see~\autoref{tab: live streaming} in~\autoref{appendix}).}

\subsubsection{Observation Data Analysis}

The researchers conducted two rounds of observation. In the first round, the three researchers each observed 30 live-streaming sessions and 30 short videos. To ensure that we cover a wide range of legal content on selected platforms, the researchers met to discuss the categories of different live-streaming sessions and short videos after the first round of observation. 

We found that the form of conducting legal live-streaming sessions is relatively fixed, i.e., live streamers hold live-streaming sessions to answer questions from the audience either verbally or literally. 
On the other hand, the legal short videos online were in various presentation forms and topics. We categorized them into different types based on the purpose of making the videos, and further identified various presentation forms within each type. 

After reaching a consensus on categorizing live streaming and short videos, the three researchers each conducted a second round of observation by sampling 20 more live-streaming sessions and 20 more short videos, following the same scheme to identify potential new content categories. After discussing the results of the new content observed, the researchers all agreed that they were aligned with the existing categorization. In summary, we observed 150 live-streaming sessions and 150 short videos from Douyin and Bilibili. The observation deepened our understanding of the legal information-seeking process on VSPs and informed the design of the question for the subsequent interviews. The observation notes were combined with the interview transcripts for an open coding analysis~\cite{corbin2015basics}, with a detailed explanation in Section~\ref{data analysis}.

\subsection{Semi-Structured Interviews with Legal Information Seekers}

We conducted semi-structured interviews with 20 legal information seekers to gain a direct understanding of their behavior when using VSPs to acquire legal information. The details are presented as follows.

\subsubsection{Interviewee Recruitment}

We recruited interviewees by i) watching legal live-streaming sessions and directly sending invitations to the information seekers on Douyin and Bilibili, ii) publishing recruitment posts on Xiaohongshu (RedNote). For the first recruitment approach, by targeting the information seekers in legal live-streaming sessions, we ensured that the potential interviewees had experience seeking legal advice from legal content creators. We used the direct messaging functions on Douyin and Bilibili to send our invitations to the targeted information seekers, in which we explained our intentions. If they agreed to participate in the study, we added them as WeChat contacts and scheduled the following interview with them on WeChat. For the second approach, we recruited interviewees by publishing recruitment posts on Xiaohongshu with the authors' personal accounts. Within the posts, we illustrated our research goals and requirements for interviewees. We encouraged viewers to contact us through the direct messaging function on Xiaohongshu if they wanted to participate. Once an information seeker contacted us, we asked them to provide some details about their experience seeking legal information by joining live-streaming consultations or watching legal short videos. We ensured that every interviewee in our study fulfilled these requirements.

A total of 20 interviewees (11 female and 9 male, all self-identified) participated in our study, aged 19 to 30. We referred to them as P1-P20 (~\autoref{tab:demographics}). The 
interviewees consumed a wide range of legal topics online, with 9 of them indicating that they were interested in all kinds of legal topics. For those who were only watching content in a certain legal field, they were most concerned about employment (5 people) and marriage (4 people). Although we did not intentionally recruit participants with different levels of legal education backgrounds, 5 participants had undergone formal legal education among the 20 people, providing diverse perspectives and insights into understanding the influence of personal legal literacy on their information-seeking behavior. Each participant received 50 Chinese Yuan (7.05 USD) as compensation for their time and contributions to the study. \zyw{Aligned with recommendations by Creswell and Guetterman~\cite{creswell2024educational}, the authors determined that 20 interviews would provide a manageable yet comprehensive dataset, allowing for an in-depth exploration of the research questions without overwhelming the analysis process.}

\begin{table}[t]
\centering
\caption{Demographic Information of Interview Participants}
\resizebox{\textwidth}{!}{\begin{tabular}{ccccccc}
\hline
\textbf{ID} & \textbf{Age} & \textbf{Gender} & \textbf{Occupation} & \textbf{Platform(s)}$^a$  & \textbf{Legal Education Background}$^b$ \\ 
\hline
P1 & 25 & Female & Legal Specialist& Douyin  & Master's Degree in Law \\ 
P2 & 23 & Female & Student & Douyin, Bilibili & -  \\ 
P3 & 21 & Male & Student & Douyin, Bilibili & - \\ 
P4 & 23 & Male & Customer Service & Douyin, Bilibili & - \\ 
P5 & 30 & Female & HR & Douyin & - \\ 
P6 & 19 & Male & Student & Douyin &  UG Student in Law  \\ 
P7 & 21 & Male & Student & Douyin  & UG student with Minor in Law \\ 
P8 & 26 & Female & Clerk & Douyin  & -  \\ 
P9 & 22 & Female & Student & Douyin, Bilibili & -  \\ 
P10 & 21 & Male & Student & Douyin, Kuaishou, Bilibili & -  \\ 
P11 & 23 & Female & Student & Douyin & - \\ 
P12 & 19 & Male & Student & Douyin, Bilibili & UG Student in Law \\ 
P13 & 22 & Female & Student & Douyin & - \\ 
P14 & 22 & Female & Student & Douyin & - \\ 
P15 & 20 & Female & Student & Douyin & - \\ 
{P16} & {21} & {Male} & {Student} & {Bilibili} & {-} \\
{P17} & {19} & {Female} & {Student} & {Douyin} & {UG Student with Minor in Law} \\
{P18} & {23} & {Female} & {Receptionist} & {Douyin} & {-}  \\
{P19} & {24} & {Male} & {Printer} & {Douyin} & {-}  \\
{P20} & {24} & {Male} & {Customer Service} & {Douyin} & {-} \\

\hline 
\end{tabular}}
\resizebox{\textwidth}{!}{\begin{tabular}{l}
$^a$Platform(s) refer to the video-sharing platforms in China that our participants used to consume legal content. \\
$^b$This is used to indicate the level of legal education participants received. ``UG'' stands for undergraduate. ``-'' means the \\participant did not receive any academic education in law.
\end{tabular}}
\label{tab:demographics}
\end{table}

\subsubsection{Interview Protocol}

The semi-structured interviews were conducted remotely through Tencent Meeting, lasting about 40-50 minutes. The question design followed the order of our research questions. All interviews were conducted in Mandarin and were transcribed into Chinese automatically via Tencent Meeting for subsequent analysis.

During the interviews, we encouraged information seekers to recall their experiences seeking legal information to understand their motivation. They were asked to compare seeking through legal content creators with other channels they knew, such as traditional offline consultations or via search engines. They were also prompted to discuss their preferred platforms and types of content, as well as their favorite legal content creators' characteristics. To understand more details about their interactions, we asked participants to recall how they interacted with the legal content creators and other audiences, what functions of video platforms they liked, and why. 

To assess viewers' considerations in choosing legal content creators, we asked them what factors affected their trust in the legal content creators concerning the reliability of their legal advice. The questions included how they searched for legal content online, their first impressions of the creators, what information they would check to make sure the creators were trustworthy, and what attributes of the creators made them feel that they were professional. The participants were further encouraged to discuss the characteristics of legal content creators they felt mistrustful of. 

Additionally, the interviews delved into understanding viewers' evaluation of content shared by legal content creators. Participants were asked how they assessed the credibility of the content, including the level of detail in legal explanations and the thoroughness of case discussions. Participants were encouraged to recall memorable cases, particularly when they felt the creator’s answers were improper or inapplicable, and to discuss how they addressed such situations when the information provided did not meet their expectations. We explored what other channels viewers would refer to when evaluating the credibility of the information they received.

\subsubsection{Interview Data Analysis} \label{data analysis}

The data collected for analysis included (1) observation notes, (2) screenshots of live streaming sessions and short videos in observation, and (3) interview transcripts, as described in previous sections. The analysis of data followed an open coding method~\cite{corbin2015basics}. Following the work of McDonald et al.~\cite{mcdonald2019reliability}, Guo and Divine~\cite{freeman2021body}, we did not seek inter-coder reliability in our analysis since our analytical procedures focused on generating concepts and themes rather than reaching agreements on the codes. Previous work indicated that even if coders agree on codes, they may interpret the meanings differently~\cite{mcdonald2019reliability, freeman2021body}. Therefore, we sought to identify emergent themes, explore the interrelationships between them, and organize them into more complex and broader themes in our analysis.

The first and second authors of the paper, who were native Chinese speakers, coded the observation notes and screenshots of the observed legal live streaming sessions and short videos, as well as the first four interview transcripts (including two male and two female interviewees) individually, and then met to discuss the codes.
Then, the first author coded all the remaining transcripts. The codes were organized under the three research questions of the paper: legal information seeking behaviors, social interactions within the legal community, and information seekers' personal evaluation of the legal information. Then, through iterative meetings among the research team members, we generated higher-level themes from the codes. Finally, we grouped the themes under common topics and translated all the themes into English for presentation in the findings.
\section{FINDINGS}

In this section, we describe three major themes that characterize legal information seeking on VSPs. In Section 4.1, we comprehensively elaborated on viewers’ legal information-seeking practice on VSPs, including livestreaming-mediated consultations and consumption of various types of legal videos (RQ1). Then, in Section 4.2, we revealed information seekers’ social interactions with legal content creators and trust factors within, as well as a description of the formation of legal knowledge-sharing communities and peer support on VSPs (RQ2). Finally, we showed how information seekers evaluate the information quality and the associated challenges they faced in Section 4.3 (RQ3).

\begin{figure}[t]
\includegraphics[width=\linewidth]{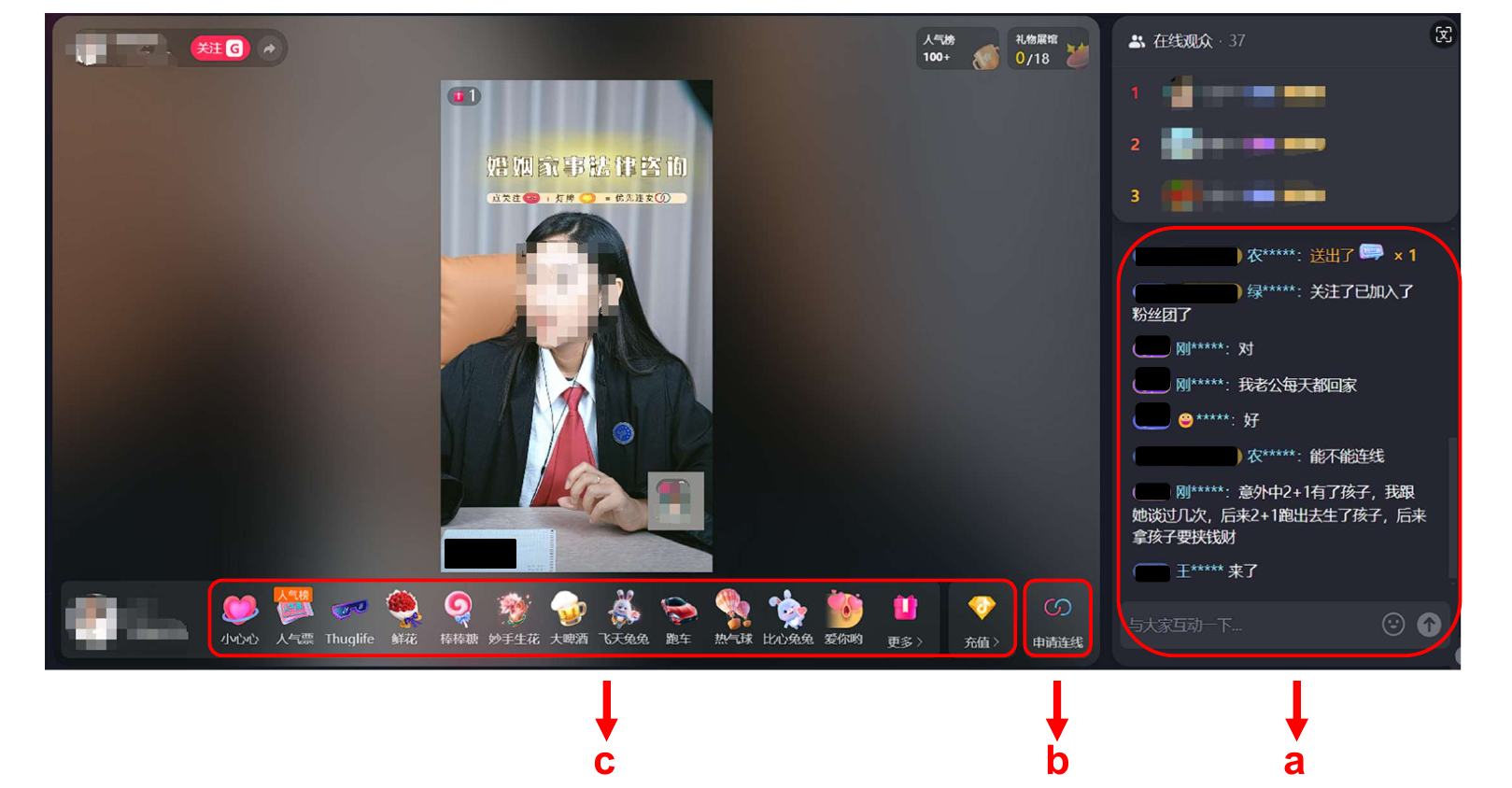}
\centering
\caption[Caption]{
The interface of a legal live streaming channel on Douyin, which includes (a) the comment (or chat) section where audiences can post real-time text comments; (b) the ``apply for connection'' button, which enables audiences to connect with the streamer and interact with them via voice, one at a time; (c) gifting functions, for purchasing and sending virtual gifts to the streamer, which also serve as the payment for consultations.
}
\label{fig:ls-screenshot}
\Description{This figure is fully described in the text.}
\end{figure}

\subsection{Legal Information Seeking Behaviors on VSPs}
\label{sec_seeking_behavior}
\zyw{This section presents our findings on how users utilize VSP affordances to fulfill their legal needs. We found that legal information seekers not only engage in interactive
and synchronous livestreaming talks with legal live streamers for deep consultations, but also consume and compare various types of legal videos to fulfill their diverse needs in daily lives. }

\subsubsection{Livestreaming-mediated Legal Information Seeking}

\label{sec-livestreaming}
Legal information seekers effectively utilized the live streaming affordances on VSPs to acquire the information they needed, including conducting one-on-one live consultations through live sessions. We presented a thorough description of this newly identified legal information-seeking behavior in this section.

\textbf{Audio-based 1-on-1 live consultations.} The most popular form of conducting legal live-streaming was similar to 1-on-1 consultations with lawyers online. The legal live streamers would invite the audience who needed advice on legal cases to phone in their live streaming sessions and raise questions through the built-in function on the live-streaming interface, as shown in section (b) of ~\autoref{fig:ls-screenshot}. Within such 1-on-1 consultations, information seekers could directly discuss their legal cases with the streamers. The streamers would usually spend 5-10 minutes listening to the information seekers' narration and sometimes interrupt to ask questions regarding specific details. For example, when hearing about cases involving fraud, the streamers would pay extra attention to the amount of money the information seekers lost or received, depending on whether the information seekers were victims or suspects. After learning about the situation, streamers would typically provide suggestions to information seekers on the next course of action to take. 

Our interviewees indicated that, compared to making appointments with lawyers offline, they preferred online consultations because they saved a great deal of effort and were more affordable. As our interviewee P4 articulated, 

\begin{quote}
    \textit{``Simply watching live streams is free of charge, and as far as I know, consulting them (legal live streamers) through live streaming is much cheaper compared with the starting price of 300 yuan (28 USD) offline. Plus, I don't have to go out!''} (P4)
\end{quote}

Furthermore, such live streams provide users with the freedom to receive detailed and tailored answers through voice communication with legal streamers, and they can even select specific presentation styles based on their preferences. As P20 described his experience in an online consultation,  

\begin{quote}
    \textit{``Perhaps because he (the legal streamer) was in front of the camera, he showed great patience, listened carefully to what I said, and further guided me on what I should do next. His responses were comprehensive compared to the answers I learned in some online forums.''} (P20)
\end{quote}

\textbf{Paying for Consultations via Gifting.} For the top legal streamers on video platforms, the cost for one counsel session was about 100-200 Chinese Yuan (14-28 USD). \cjl{The entire payment process was conducted through the video-sharing platform's gifting system.} For example, in Douyin, 10 ``diamonds'' (one of the virtual gifts on Douyin) cost 1 Chinese Yuan, and the transaction with the streamers is done by purchasing such virtual gifts. Some lesser-known creators may offer free sessions to attract information seekers. Each session lasted about 10-20 minutes, depending on the complexity of specific legal cases. 

\textbf{Live Q\&A Sessions.} Besides audio-based consultations, some users chose to raise their queries in the live comments (similar to chat in Western live-streaming platforms like Twitch), which was a built-in function in the platform for audiences to post text comments, as shown in~\autoref{fig:ls-screenshot} (a). Such legal content creators were usually less famous and received fewer queries during the live-streaming sessions, which allowed them to review the questions in comments and provide more detailed suggestions. Such a text-based live-streaming consultation was mostly free of charge and also attracted some information seekers. For example, P11 reflected in interviews that she preferred to type the questions out rather than directly talking to the creators because of her introversion.

\subsubsection{Video-based Legal Information Seeking}
\label{sec: short videos}

Legal information seekers also consumed various legal videos on VSPs to acquire the necessary legal knowledge. Our observation and interviews identified three significant forms of legal short videos: videos promoting legal knowledge, legal course recordings, and archived legal live streams. 

\textbf{Promoting Legal Knowledge.} The first type encompasses the various short videos that promote legal knowledge. These videos were presented differently, according to the creators' style, including direct legal instructions, legal scenarios performed to explain the legitimacy of certain behaviors, or simply showing the text descriptions of the law. Such videos were typically one to two minutes long, focusing on a specific legal case. The interviewees had different preferences for such videos that promote legal knowledge. For instance, P12 expressed his fondness for videos that performed legal scenarios:

\begin{quote}
    \textit{``I probably prefer legal scenarios performing videos because it tells a real case and then combines legal knowledge with it. First of all, it is easier to understand. Another reason is that by watching such videos, we can better understand some of these situations and help ourselves analyze similar cases in reality.''}
\end{quote}

On the contrary, P5 liked the videos that directly listed legal information in words, and she explained, 

\begin{quote}
    \textit{``I prefer the videos with many words and explain things in detail because they contain more useful information (compared to other videos)''}.
\end{quote}

\zyw{Some participants further expressed that they liked the richness of video content on VSPs. For example, P5 recalled her experience searching for how to protect her property in a divorce and stated, \textit{``This (divorce) is a severe issue for me, so I cannot easily decide on my next move. I can quickly find content from different lawyers online by searching with the keywords `property protection in divorces' (on video platforms). I can easily compare their opinions and evaluate what is best for me to do now.''} The abundance of content on VSPs enabled viewers to compare different videos quickly. }


\textbf{Recordings of Legal Courses.} In addition to the short videos that promoted legal knowledge, some recordings of legal courses were also popular among information seekers. The content typically focused on educating legal students to help them prepare for professional exams. The presenters in such videos were mostly legal scholars or university professors. Such course recordings on video-sharing platforms were usually longer than other short videos, lasting about 5-10 minutes. Many were drawn to the presenters' engaging teaching styles. \zyw{For example, one famous legal content creator, \textit{``Luo Xiang Talks about Criminal Law''} (\begin{CJK*}{UTF8}{gbsn}罗翔说刑法\end{CJK*}), was known for his series of videos that explained questions in the Chinese legal professional exam. 
He has attracted over 30 million followers on Bilibili, including law students and a wide range of ordinary viewers.}

\textbf{Archived Live Streams.} The last type of video was archived live streams, mainly replaying some 1-on-1 consultations with legal content creators. Some interviewees noted that they regularly consumed such archives because they were unable to catch the live-streaming sessions and preferred to watch the videos in their spare time. Creators often edit archived live streams to reduce idle moments, making these video archives more efficient for information seeking. As P13 explained, 

\begin{quote}
    \textit{``I prefer such archived live streams because the creators will usually choose some classic cases and make a conclusive title of them. Therefore, I can easily select the ones I am interested in.''}
\end{quote}


\subsection{Legal Interactions and Formation of Trust on VSPs}
\zyw{In this section, we present how information seekers' interactions with legal content creators and peers shape the formation of trust on VSPs. }

\subsubsection{Interactions and Trust with Legal Content Creators} 

\zyw{We found that legal information seekers selected and established trust in content creators based on the impression created by their presentation styles, and reasons for further engagement included reduced epistemic discomfort and empathetic emotional support from the creators.} 

\label{trust_with_creators}

\textbf{Trust to the First Impression of Legal Content Creators.} 
We found that viewers' first impression of legal content creators affects their perception of their professionalism, including a fluent and firm tone, a good personal appearance, and impartial attitudes toward information seekers. Legal content creators used such professional self-presentation to gain viewers' trust and admiration for their expertise.

Many participants preferred creators who could speak fluently and quickly "catch the main points". For instance, P3 stated that the fluency of creators made him trust their expertise and described one of the creators he followed as \textit{``She (the creator) was able to answer the audience's questions immediately or quickly come up with solutions to solve issues in certain scenarios. Then I didn't see any problem with her professionalism.''} Many participants admired the creators' ability to ``see through'' what the audiences hid behind. They reflected that information seekers often deliberately hid some details they were ashamed of or ignored important information while stating the issues, and they found it enjoyable to watch the process of live streamers digging out the truth. As P15 illustrated, 
\begin{quote}
    \textit{``Maybe the person (information seeker) did not want to say some unfavorable information about himself, but the lawyer (legal content creator) was able to use his eloquence to extract the information he wanted, which I think is very admirable. After the counseling, he would also educate the audiences about the law in simple words, not as difficult to understand as the written cases, so that I could understand easily''} (P15)
\end{quote}


Besides the eloquence, a good personal appearance could also increase viewers' trust in the creators. As P4 described, the creators who \textit{``wear glasses and a suit, look like a foreign ministry spokesman''} gave him a reliable feeling. Moreover, some participants appreciated the impartial attitudes that legal content creators displayed towards their audiences, whether they were victims or suspects, and considered it a professional performance. P13 praised one creator as \textit{``always standing in an objective position, and would not defend you because you are potential clients''}. She liked the creators for always educating their audiences on the right behavior. 

\textbf{Humor and Friendliness Reduce Viewers’ Epistemic Discomfort.} Many interviewees reflected that the content created by legal content creators helped mitigate their epistemic discomfort in talking to legal specialists and broke their stereotypes of lawyers being too serious or even ``scary''. 

Our participants expressed a preference for a casual and relaxing communication atmosphere within live streams and short videos featuring legal content creators. As legal content creators often presented themselves in a friendly or humorous manner, viewers felt more at ease interacting with them online. One female interviewee, P8, stated her stereotypical feelings about traditional lawyers, e.g.,

\begin{quote}
    \textit{``I sometimes have a stereotypical image of lawyers being too strict and fierce, and I might be a bit scared of them. Perhaps it is because I have watched too many Hong Kong TV dramas; lawyers are often portrayed as mean and scary. So I will definitely prefer to talk to them remotely''} (P8)
\end{quote}

Similarly, P4 expressed his feelings about the arrogance of lawyers offline and stated, \textit{``I think it would also be less of a rushed feeling online. There are some slightly arrogant lawyers offline, maybe only a small percentage, but it will affect how I feel (when talking to them).''} Many participants felt nervous or shy to talk about their personal issues face-to-face with lawyers. Therefore, online consultations that do not require users to show their faces can offer an ideal space for them to elaborate on their concerns in a more relaxed manner.

When watching legal videos, the participants preferred those explaining legal cases in a humorous way or delivering legal knowledge in plain language comprehensively to people without legal expertise, such as \textit{``Luo Xiang Talks about Criminal Law''} (\begin{CJK*}{UTF8}{gbsn}罗翔说刑法\end{CJK*}), who was introduced in detail in Section~\ref{sec: short videos}. Many participants reflected that they were unaware of the existence of legal content creators until they were drawn to Luo Xiang's humorous and comprehensible video style.

\textbf{Female Viewers Value Empathy from Creators.} Among the 11 female interviewees, five (P2, P5, P8, P9, P15) mentioned that they preferred to turn to female creators because they sensed empathy as women from the creators. They felt that female creators could better stand in women's point of view and provide advice and emotional support to female information seekers. One frequently mentioned female creator was \textit{Lawyer Long Fei} (\begin{CJK*}{UTF8}{gbsn}龙飞律师\end{CJK*}), who specialized in answering legal issues on marriage and love topics. Among her 9 million followers on Douyin, more than 70 percent were young women between 24 and 30, according to a report on her Douyin account. As one female interviewee summarized her observation of lawyer Long Fei's channel,

\begin{quote}
   \textit{``On lawyer Long Fei's channel, the information seekers are basically women who want to consult about some emotional problems (particularly relationship issues). For example, if they found their boyfriends were cheating or buying sexual services, what would happen? During their consultations, they will talk more about their emotional considerations. ...... Long Fei will not only give some legal advice but also provide them with some emotional support.''} (P2)
\end{quote}

\subsubsection{Interactions and Long-term Engagement within the Peers' Community}
\label{sec_peer_community}
\wzy{Compared with other information-seeking channels like search engines or legal websites, the performative nature of live-streaming and short videos makes such information-seeking behavior involve more than the seekers and responders. For example, even within the 1-on-1 live-streaming consultations, other viewers watch on the same channel. Some interviewees reflected that they received emotional support from other viewers and even constructed a long-term engagement with the follower community to learn and share information. }

For example, P5 described her interactions with other viewers who shared similar experiences on live streaming channels or in video comments, and she can also learn from others' stories and receive emotional support. She stated, \textit{``When I watched live streams or videos, I realized that there were many women who had been in the same situation as me (husband cheating), and they shared their stories in the comments section. When I commented, they would also offer me emotional support, making me feel like I was not alone.''}

We found that several interviewees established long-term engagement with other viewers by joining fan groups of certain legal content creators, either within the video-sharing platforms or on instant messaging apps such as WeChat. Two of our participants (P11 and P12) stated that they had joined the fan groups of creators on WeChat. \cjl{We closely observed how legal content creators promoted their WeChat fan groups during live-streaming sessions on Douyin. Typically, a link to join the group is provided in the live-streaming interface. Still, there are often specific requirements for audiences to participate, such as following the creators' accounts or paying by sending a certain number of virtual gifts on Douyin.}
P12 mentioned that the fan group he was in consisted of about 30 members, and he joined it out of convenience to learn some legal information from time to time and to contact the creators when needed. P11 described her experience in the fan group in more detail, stating that joining the WeChat group provided her with timely legal knowledge and entertainment.

\begin{quote}
    ``\textit{There were dozens of followers in the WeChat group. We interacted a lot, sometimes using voice group chat and sometimes listening to others who shared their cases, while the creator himself would occasionally share some legal knowledge with us. I'm on there mainly to listen to them share stories and build up some knowledge.''} (P12)
\end{quote}

Such social interactions within fan groups facilitated long-term engagement among followers. This approach enabled them to receive timely feedback from others on legal issues and facilitated continuous self-improvement, motivating some of our participants to maintain long-lasting connections with the follower community.

\subsection{Personal Evaluation of Legal Information Quality on VSPs}
\label{sec_evaluation}
\zyw{In this section, we present our findings on information seekers' evaluation of legal information quality on VSPs. We found that compared with people with higher legal literacy, laypeople lacked the awareness, willingness, and cross-validation approaches to verify legal information. }

\subsubsection{Personal Legal Literacy Affects Awareness and Judgments of Content Quality}
\label{sec_personal_literacy}

In this study, we found that the interviewees had different levels of legal education backgrounds, which provided us with insights into how they perceive the content quality of legal content creators differently. 

Among our 20 participants, 5 (P1, P6, P7, P12, P17) were already holding or pursuing a law degree. In contrast, others had not received any systematic legal education or training and can be considered laypeople. When we asked them about their evaluation of the information quality they received from legal content creators, we found that their legal literacy could affect their judgments on the effectiveness and applicability of legal information. 

When asked whether they sometimes felt that the legal information or advice delivered by legal content creators lacked integrity or applicability, most laypeople in our interviews responded that they had never noticed any problems regarding the quality of the information. In contrast, interviewees with higher legal literacy showed concerns about the low-quality content created by creators. For example, an undergraduate student in law (P7) indicated, 

\begin{quote}
    \textit{``I think that the quality of content highly depends on the creators themselves. Even for the same creator, I can not say that all of his work captures my interest. Sometimes, I don't agree with his analysis of certain cases. For the same creators, there are some videos that I will watch, but there are others that I may just stop watching halfway through.''} (P7)
\end{quote}

P1, who was a legal specialist, further shared her concerns about the information's applicability, 

\begin{quote}
    \textit{``According to our profession, the legal issues can be divided into two: one is the factual, and the other is the legal problem. In fact, such online consultations are more aimed at helping the audience resolve a factual level of a problem. Regarding practice, the situation will be much more complicated, and I don't think some of the advice is applicable in reality.''} (P1)
\end{quote}

Compared to legal professionals, laypeople cannot properly evaluate the integrity or applicability of the knowledge and advice they acquire from creators on their own, which may cause trouble in their lives. For example, P2 shared her experience in seeking advice from legal content creators about child custody after divorce, 

\begin{quote}
    \textit{``The creator only told me the first step was to get the husband to write a guarantee, but didn't tell me what I should do if he refused. I did what he said, but the husband didn't cooperate, so I didn't know how to proceed.''} (P2)
\end{quote}

These examples suggest that the advice or knowledge provided by creators may be lacking in practical applicability, although it may be theoretically correct. However, information seekers with lower legal literacy may not be able to properly evaluate whether such legal suggestions are suitable for their situations.

\subsubsection{Cross-validation with Other Information Channels}
\label{other_channels}
Some interviewees would evaluate the quality of content based on information received from other channels, including other social media platforms, search engines, reading materials, consulting legal professionals, and even seeking help from generative AI. 

We found that information seekers with a higher level of legal literacy tended to consult reading materials and seek advice from legal professionals. For example, P12, who was an undergraduate student in law, summarized,

\begin{quote}
    \textit{``(To verify the information I received) I would go to some websites that contain judgments from the court and then search for relevant cases. If there is no relevant information, I may ask my classmates or professors for help in solving the problem. If I can't, I may consider consulting a lawyer.''} (P12)
\end{quote}

Similarly, P1, P6, and P7 indicated they would refer to some legal professionals they had contact with, such as colleagues, professors, and peer students, for suggestions or to read some legal materials first. However, most of the other interviewees could not easily access legal specialists to help them evaluate the information. In addition to that, they had difficulty comprehending official legal materials, so they would continue to search on video platforms like Douyin and Bilibili (P3, P10, P15) or refer to other social media sites like Xiaohongshu (P2, P8, P9, P15) or search engines like Baidu (P2, P4, P8, P15). Even if they found multiple versions of information online for comparison, sometimes they still could not easily decide on which action was best for them. It remained challenging for information seekers with lower legal literacy to comprehend and apply the information from legal content creators.

Another cross-validation approach was using generative AI. We found that the interviewees expected more applications of Generative AI in facilitating legal information evaluation, but also expressed concerns about the current drawbacks of AI that made them non-competitive compared with human lawyers. 

All interviewees agreed that generative AI could potentially help them verify the information they received from legal content creators, and they were willing to try if the accuracy and integrity of advice or knowledge provided by AI could be ensured. Yet, they were concerned about whether AI could provide emotional care to humans during consultations and whether the information provided was complete and up-to-date. For instance, P13 stated his opinions on AI based on his impression, 

\begin{quote}
    \textit{``AI is a machine after all, and it cannot take an emotional standpoint. If I communicate with a real lawyer, they can stand in my position to defend me and help me analyze the situation. But all AI can do is provide me with some specifics of the laws and regulations.''} (P13)
\end{quote}

Another interviewee stated that she had little faith in the accuracy of information generated by AI, as she lacked confidence in the prompts she wrote. 

\begin{quote}
    \textit{``When I'm describing things that are complex to AI, I don't feel like doing a good job of giving the AI the correct prompts that give me the response I expect.''} (P14)
\end{quote}

These reflections revealed that AI could potentially be a useful tool in helping legal information seekers evaluate the information from VSPs, but users were still concerned with the accuracy and integrity of generative AI information due to difficulties in comprehending the rationale of how AI works, what it is capable of, and how to correctly interact with it.  

\section{DISCUSSION}



As an important mission in legal empowerment~\cite{golub2010legal, van2008access} and public legal education~\cite{PLEDefinition}, effectively facilitating the legal information-seeking behavior of the general public has been widely discussed in academia~\cite{maggs2006free, denvir2016online}. 
Our findings suggest that the various affordances of VSPs effectively support users' information-seeking behavior, even forming a unique ecosystem for legal information and knowledge sharing that involves multiple stakeholders. However, there were still risks concerning the quality and accuracy of the information, which limited the effectiveness and real-life impact of such information-seeking behaviors. \zhiyang{In this section, we discuss the potential for VSPs in transforming people's legal information-seeking behavior (section~\ref{potential-for-vsp}), risks of low-quality information and heuristic thinking (section~\ref{risk-information}), and benefits and potential biases within the VSP-mediated legal communities (section~\ref{discuss_community}). Finally, we shed light on the design implications in section~\ref{design-implications}.}

\subsection{Lowering Barriers and Sustaining Engagement in Online Legal Information Seeking}
\label{potential-for-vsp}
This work demonstrates the potential for VSPs in transforming people's legal information-seeking behavior through their various affordances, making it more approachable and engaging. The livestreaming-mediated consultations naturally exploited the advantages of synchronicity and interactivity of livestreaming~\cite{he2023seeking}, significantly lowering the barriers of lawyer-client interactions compared with traditional approaches, such as legal texts or formal consultations, which can be intimidating or overly technical. In addition to live streams, short videos, despite their brief duration, serve as effective entry points to legal knowledge.

As introduced in Section~\ref{sec-livestreaming}, the information seekers reflected that the existence of live legal consultations made it more affordable for them to access legal experts and receive professional legal assistance. The ``performative'' nature of livestreaming makes such live sessions easier to engage a wider audience and is naturally in line with the broader goal of legal empowerment in raising public legal awareness~\cite{PLEDefinition}. Moreover, we also identified that the direct interactions with legal content creators on VSPs helped the information seekers to break the stereotypical images of legal professionals being mean or even ``scary''~\cite{marchetti2008dispelling}, and reduced epistemic discomfort often associated with legal jargon and complex procedures.
Many reflected that they were more willing to engage with legal content creators or learn from them after watching their content on VSPs, as detailed in Section~\ref{trust_with_creators}.
The easy-going or humorous presentation styles of legal content creators demystify complex legal concepts, encouraging information seekers to explore topics they might otherwise avoid, especially those with a limited prior legal background.

Moreover, our findings highlighted that short videos served as effective ``gateways'' to rich legal knowledge for diverse information seekers, due to their concise nature, which aligns with the shrinking attention spans of VSP users~\cite{opara2025impact}. 
What is promising is that these short videos are delivered in an engaging manner, often leveraging humor and real-life scenarios that capture viewers' interest and motivate them to explore further. Viewers who initially engage with short legal videos may start consuming longer-form legal content over time, such as livestreaming consultations, as their curiosity and confidence in legal knowledge grow. The interactive nature of livestreaming further sustains engagement, fostering a sense of collective learning within the community. These findings highlight the benefits and a promising future for VSP-mediated legal information seeking.

\subsection{Risks to Information Seeking: Overly Relying on Heuristic Processing}
\label{risk-information}
Despite the various benefits brought by the affordances of VSPs, our findings also uncover the potential harm of low-quality, or even incorrect legal information, which was rooted in the performative and fast-paced nature of VSPs, and their tendency to prioritize \textit{heuristic processing} over \textit{systematic processing} \cite{chaiken2012theory}.
Short videos and livestreams excel at capturing viewers' attention through affective cues, such as humorous presentation styles and emotional storytelling, which encourage rapid, heuristic-based judgments among viewers.
For example, our findings suggest that to ensure the smoothness of livestreaming sessions, the streamers tended to make rushed conclusions and suggestions on the legal queries raised by information seekers. 
As introduced in Section~\ref{sec-livestreaming}, each consultation typically only lasted around 10 to 20 minutes, and many participants reflected that the streamers did not delve too deeply into details before making their suggestions. 
Our findings suggest that legal information seeking on VSPs encourages heuristic information processing, as it focuses more on obtaining a quick answer rather than rigorous analysis of the reasoning, which poses significant risks to the quality and accuracy of the legal information or advice provided by content creators.
This heuristic dominance aligns with the cognitive ease of video consumption on VSPs, as viewers can absorb information passively without the effortful scrutiny required by textual or formal legal sources.



However, legal understanding and sense-making often demand deeper \textit{systematic processing}, where individuals critically evaluate arguments, verify sources, and weigh contextual nuances.
\zhiyang{Prior studies have shown that laypeople often cannot properly evaluate legal suggestions and tend to consider them as concrete results due to a lack of specialized domain knowledge~\cite{dewan2024teen, zhang2022shifting, thomas2014online}. Similarly, Dewan et al.~\cite{dewan2024teen} reported that teens in the U.S. could not easily identify misinformation when seeking information about reproductive health on social media. Our study aligns with these studies, demonstrating that the ability of legal information seekers to correctly comprehend and make sense of the legal information shared on VSPs may depend on their personal literacy (see Section~\ref{sec_personal_literacy}). Many interviewees, especially those with limited legal literacy, were unaware of the importance of evaluating information in online legal information seeking.} We further identified that the professional presentation styles adopted by many legal content creators could negatively affect viewers' critical thinking on their answers. As presented in Section~\ref{trust_with_creators}, viewers could be easily engaged by the eloquence and professional appearance of legal content creators and 
accept their legal opinions without question. 
The very features that make VSPs engaging, including brevity, entertainment value, and algorithmic recommendations, may inadvertently suppress systematic processing. Cognitive load of video consumption further compounds the issue because the multitasking nature of VSP use drains mental resources that could otherwise be allocated to support reflective and critical thinking.
This tension raises a critical design and ethical question: How can VSP-based legal information seeking strike a balance between heuristic processing and the systematic processing required for robust legal understanding and sense-making? 

Therefore, despite the promises of legal information-seeking through VSPs, we call for more investigations into mitigating the risks of inaccurate and low-quality information as well as promoting viewers' critical thinking.

\subsection{Navigating the Nature, Promises and Risks of VSP-Mediated Legal Communities}
\label{discuss_community}

Prior research on information seeking on social media revealed that the various social media platforms have become one of the primary channels for users to seek advice on important issues related to their daily lives or self-development, including healthcare~\cite{augustaitis2021online,dewan2024teen,milton2024seeking,starling2016information, zhang2022shifting}, career development~\cite{blaising2019navigating, tomprou2019career, malik2024towards} and other important areas~\cite{hirsh2012reality,jonas2024we,nicholson2019if}. 
\zhiyang{In this section, we situate our findings within prior literature on online communities for knowledge-sharing, highlighting the unique characteristics and promises of VSP-mediated legal knowledge-sharing communities, but also emphasizing the associated risks.} 


\zhiyang{In our study, many interviewees had no previous connection in their daily lives with legal specialists like lawyers or judges, as reflected in Section~\ref{sec_personal_literacy}.
Their lack of access to domain experts for consultations in their social circles reduced their willingness and confidence in referring to other information sources to cross-validate legal information shared on VSPs.  Some even demonstrated intimidation in reading legal materials on their own. However, the affordances of VSPs, the production quality of online legal content, and the opportunities to directly interact with legal content creators and engage with communities of other information seekers make it easier for users to seek legal information through VSPs.
Therefore, our results highlighted the significance of online knowledge-sharing communities in supporting laypeople's learning and evaluation of specialized domain knowledge shared on VSPs. As presented in Section~\ref{sec_peer_community}, the naturally-developed VSP-mediated communities were particularly valued by the interviewees, serving as irreplaceable resources for continuous learning, information sharing, peer emotional and social support, and cautionary tales. Sharing legal information on VSPs not only promotes information seekers' understanding of specialized domain knowledge but also enhances their awareness of information evaluation and critical thinking. 
These findings resonate with recent HCI works that have uncovered the value of many other communities on social media in supporting the members for diverse purposes, such as senior matchmaking~\cite{he2023seeking},
health-related discussion~\cite{harrington2019engaging}, and coping with cybersecurity and privacy
threats~\cite{kropczynski2021towards, murthy2021individually}. 
However, we noticed that only two interviewees in our study reflected actively engaging with such communities, which poses a critical question for future research on how to naturally cultivate communities on VSPs to facilitate information-seeking while promoting peer support and social engagement.}

\zhiyang{Despite the benefits brought by such online communities for sharing legal information, it is also important to note that they are still susceptible to biased information or even misinformation. Firstly, we warn of the potential risks of echo chambers in these communities, the self-reinforcing environments where users primarily encounter information that reinforces their existing views~\cite{yu2025breaking}. \zyw{Past research has documented the spreading of misinformation in many online communities due to the effect of echo chambers~\cite{ghafouri2024transformer, guess2020exposure}, which could have even affected the results of the 2016 US presidential election~\cite{guess2020exposure}. }
In our study, the members in VSP-mediated legal communities were often described by our interviewees as like-minded individuals who shared similar values. Therefore, the members were unlikely to be exposed to diverse perspectives on legal issues, and hence, increasing the risks of spreading biased or misinformation within the communities. Moreover, uncritical trust in influential creators within online communities may also play a significant role in fostering biases~\cite{metzger2013credibility}. Although influential creators are often legal specialists with knowledge of the law, past research has shown that even legal experts can be influenced by extrajudicial factors, leading to biased judgments in legal cases~\cite{danziger2011extraneous, 2014fluency, kapardis2016extra, rachlinski2008does}. For example, an experiment on American judges showed that emotional reactions to litigants could influence judges' decisions, and they favored those they liked or felt sympathy for, but reacted more negatively towards those they disliked and felt disgust~\cite{wistrich2014heart}. Therefore, when creators were influenced by their own preferences and made irrational judgments on certain legal cases, community members could be easily affected by this biased information due to their trust in content creators. }

\subsection{Design Implications}
\label{design-implications}
Based on our findings, we propose the following design implications to better support legal information seeking. 

\subsubsection{Design for Encouraging High-Quality Legal Content Creations}

Our findings highlighted that the quality of legal content on VSPs cannot be effectively ensured under the current platform recommendation algorithms. For instance, our observations of legal livestreaming sessions and interviewees' reflections suggested that livestreamers often provided rushed answers to information seekers to ensure smoothness and maintain audience engagement. 
Similarly, a study on the recommendation algorithms adopted by TikTok shows that TikTok favors creators who post frequently with popularly used hashtags and highlights them to more viewers~\cite{klug2021trick}, and another study reveals that the current platform design may incentivize legal content creators to prioritize quantity over quality and give poorly-researched or even inaccurate advice~\cite{mcpeak2019internet}. 
\zyw{
To mitigate the tension between the creator’s need for visibility and the user’s need for accuracy, VSPs must look beyond engagement metrics. We suggest a collective curation mechanism for legal content. 
Platforms could implement a legal-context annotation system similar to crowd-sourced fact-checking tools. In this design, users with verified legal credentials can attach legal context, corrections to legal interpretations, or warnings of inaccuracy to legal livestreaming channels or videos. This is similar to the idea of social-media scale curation introduced in the work of He et al.~\cite{he2023cura}, but we suggest crowd-sourced annotations instead of algorithmic predictions. These annotations would be visible to users as an overlay, ensuring that viewers receive additional information about content quality. Moreover, the platform's recommendation algorithm could then be retrained to weight these "expert endorsements" as a significant factor in recommendations, thereby promoting quality over quantity at a platform level over time. 

} 

\subsubsection{Design for Efficient and Reliable Legal Information Verification}
In this study, we found that it still challenged most interviewees to efficiently evaluate legal information on VSPs, \zyw{due to low legal awareness, a lack of specialized knowledge, and limited verification approaches, as introduced in Section~\ref{sec_evaluation}.} 
To this end, it is warranted for VSPs to incorporate built-in mechanisms that can promote critical thinking to the users in consuming legal content and facilitate
convenient legal information cross-validation. In a study on misinformation on YouTube, Hussein et al. found that users’ watch history creates a filter bubble, causing YouTube to continuously recommend similar misinformation to viewers~\cite{hussein2020measuring}. To prevent such cases in legal information seeking, we recommend that VSPs adopt more flexible recommendation mechanisms and expose viewers to different legal opinions, allowing them to be aware of the importance of careful information validation and to effectively compare the different advice. \zyw{Specifically, VSP designers can consider implementing diversified recommendation clusters. Instead of only promoting similar next videos, VSPs could additionally present a list of suggested videos that offer alternative legal opinions or reasoning on the same topic. From a user's perspective, this transforms the passive act of watching the next video into an active choice between different legal arguments, reminding the user that the legal content they are consuming is subjective and requires validation against other sources.}

Future VSP designers are also advised to adopt new technologies, such as AI, to support more accurate legal content moderation and provide convenient and reliable verification services for legal information on VSPs. The development of large language models (LLMs) presents new opportunities for verifying reliable legal information. For example, recent investigations on legal LLMs have revealed their potential in providing intelligent legal services and explaining complex legal concepts to non-experts~\cite{jiang2024leveraging, yue2023disc}. 
\zhiyang{However, as we found that many people had already started involving LLMs in their legal decision-making \zyw{(see Section~\ref{other_channels})}, we would like to emphasize the ethical concerns about LLM-facilitated legal information verification. As LLMs are typically trained on an enormous scale of real-world data sourced from the Internet~\cite{park2025autistic, gallegos2024bias}, studies have shown concerns about LLMs inheriting biases in their training data~\cite{bender2018data, blodgett2020language}. These biases reflect the societal prejudices, stereotypes, misrepresentations, derogatory and exclusionary language, and other denigrating behaviors in the human society that arise from historical and structural power asymmetries~\cite{park2025autistic, gallegos2024bias, guo2024bias}. In the legal field, the development and application of LLMs are associated with potential challenges of perpetuating and even amplifying the biases rooted in the LLMs~\cite{app15168860, liu2024judges}, and pose serious ethical risks to the marginalized communities by leading to unequal treatment or skewed results towards them~\cite{guo2024bias, gallegos2024bias, bender2019typology, sheng2020towards}. Therefore, we call for future legal LLM developers and practitioners to focus on the reliability and fairness of LLM-facilitated legal information verification, particularly whether LLMs could strictly follow the principles of the rule of law and disregard the influence of extrajudicial factors in the reasoning process.}



\subsubsection{Design for Supporting Legal Information Seeker Community}

In this section, we discuss potential future VSP designs that can enhance the building, maintenance, and thriving of online communities of legal information seekers. First, this study revealed that the existing communities were mostly creator-centered, meaning that they were initiated and maintained by famous content creators, and members were recruited from the creators' follower groups. \zyw{This limited the influence and visibility of such communities to a wider group of information seekers, as indicated by our findings, which show that only 2 out of 20 interviewees reported knowing and joining follower groups for some influential creators (see Section~\ref{sec_peer_community}). 
Therefore, we call for VSP designers to consider scaling up these small, creator-centered communities to platform-wide communities, in order to benefit a wider group of users with peer support and better accessibility to information. VSP designers can consider implementing topic-centric communities that aggregate legal content and discussions across multiple creators. Unlike traditional text-based legal forums, these communities can leverage the VSP affordances, for example, allowing users to post text queries, which verified creators can answer via short video replies. Moreover, as we observed that the existing creator-centered groups were often associated with certain requirements to join, such as following the creators' accounts or paying by sending virtual gifts to the creators on VSPs, the construction of such publicly available legal communities on VSPs could also help lower the barriers for regular users to enjoy the power of community support in seeking and evaluating legal information.}

\section{LIMITATIONS}

\zhiyang{This work has several limitations. Firstly, in this study, we selected two video-sharing platforms in China, Douyin and Bilibili, to observe legal live streaming sessions and legal short videos. Although we provide justifications about our selection of these two platforms in Section~\ref{plat-selection}, our sampling might not be fully representative of all VSPs in China, and our data collection might not cover all the potential forms of legal information on VSPs or comprehensively categorize all the legal topics. We call for future work to validate and cross-compare our findings by exploring a larger corpus of data in a broader range of platforms through quantitative or computational methods. Second, due to the demographic characteristics of interviewee recruitment platforms, the participants in our interview study might not fully represent all the legal information seeker groups. We notice that the interviewees in our study tend to be young (less than 30 years old) and familiar with the functioning of VSPs and even more advanced technologies like LLMs. However, it is essential to note that the senior population in China also has legal needs, but might lack the ability to use high-tech products to correctly seek and validate legal information on VSPs compared to the younger generations. Therefore, we advocate for future work to reach wider user demographics, especially the aging and less tech-savvy groups. Finally, as our study is conducted on Chinese VSPs and under the Chinese legal system, the findings might not be generalizable globally, particularly because of different legislation and regulations on lawyers offering legal advice online in different countries. We hope future work could investigate the practices of legal information seeking on VSPs under different legal contexts.}
\section{CONCLUSION}

Seeking legal information on video-sharing platforms (VSPs) has been popular among people with legal needs in recent years. Compared with traditional lawyer-client interactions, the affordances of VSPs support more convenient and affordable legal information seeking, but come with concerns about inconsistent quality and misinformation. This study makes the first investigation into the practices and challenges of legal information seeking behaviors on VSPs through observation and semi-structured interviews with 20 information seekers in China. We comprehensively revealed the current practice of VSP-mediated legal information seeking, including live legal consultations and legal short video consumption. We further reported the formation of VSP-based legal knowledge sharing communities and the social interactions within, and uncovered the influence and significance of such peer support communities in facilitating legal information seeking. We revealed the personal strategies adopted by information seekers in legal information evaluation, as well as the associated challenges. This study concludes with design implications on how to foster an accessible, safe, and efficient legal information-seeking environment on VSPs.

\bibliographystyle{ACM-Reference-Format}
\bibliography{sample-base}

\clearpage
\section{APPENDIX}
\appendix
\section{Categories of Legal Topics}
\label{appendix}

\begin{table}[h]
\caption{\zhiyang{Categories of the legal topics and example legal questions observed in live streaming sessions and short videos}}
\resizebox{\textwidth}{!}{\color{black}\begin{tabular}{l|l|l}
\hline
\textbf{Chinese Laws and Definition$^a$}                      & \textbf{Legal Topics Observed}       & \textbf{Example Legal Questions}                                                                                                                    \\ \hline
\multirow{5}{*}{\begin{tabular}[c]{@{}l@{}}Civil and Commercial Laws\\ \\ \textit{These laws cover civil rights, disputes,}\\ \textit{and commercial activities, defining}\\ \textit{the rights and obligations of individuals}\\ \textit{and legal entities in private matters.}~\cite{chineselaws} \end{tabular}} & Debt Dispute                & \textit{\begin{tabular}[c]{@{}l@{}}"How can I make the people who owe me money \\ to pay their debts?" (live stream) \\ \textit{"What methods will the court employ to deal} \\\textit{with defaulting debtors?"} (short video)\end{tabular}}                                                \\ \cline{2-3} 
                                           & Contract Fraud              & \textit{\begin{tabular}[c]{@{}l@{}}"What should I do when my boss refuses to pay \\ me the amount he previously agreed to?" \\(live stream/short video)\end{tabular}}                           \\ \cline{2-3} 
                                           & Matrimonial Assets Division & \textit{\begin{tabular}[c]{@{}l@{}}"My husband cheated on me. How can I get more \\ matrimonial assets in the divorce?" (live stream/\\short video)\end{tabular}}                               \\ \cline{2-3} 
                                           & Child Support               & \textit{\begin{tabular}[c]{@{}l@{}}"My husband refuses to pay child support after \\ the divorce. What can I do?" (live stream/short video)\end{tabular}}                                     \\ \cline{2-3} 
                                           & Shareholder's Rights        & \textit{\begin{tabular}[c]{@{}l@{}}"I was excluded from the meeting as a shareholder \\ of the company. How can I defend my rights?"\\ (live stream)\end{tabular}}                  \\ \hline
\multirow{4}{*}{\begin{tabular}[c]{@{}l@{}}Criminal Laws\\ \\ \textit{These laws focus on defining crimes,}\\ \textit{establishing penalties, and outlining the} \\ \textit{process for investigating, prosecuting,} \\ \textit{and punishing offenders. }~\cite{chineselaws} \end{tabular}}             & Fraud                       & \textit{\begin{tabular}[c]{@{}l@{}}"Can you help me analyze if I was scammed in this \\ situation? {[}...{]} What should I do next?" (live stream)\\"Be cautious of scams during the New Year celebrations."\\ (short videos)\end{tabular}}                  \\ \cline{2-3} 
                                           & Defamation                  & \textit{\begin{tabular}[c]{@{}l@{}}"I want to hold the person who defamed me online \\ accountable. How do I sue him?" (live stream)\\"How to determine whether defamation has occurred?"\\ (short video)\end{tabular}}                                \\ \cline{2-3} 
                                           & Illegal Fund-Raising        & \textit{\begin{tabular}[c]{@{}l@{}}"The owner of my company has been detained for \\ illegal fund-raising. How can he reduce the \\ sentence?" (live stream)\end{tabular}}        \\ \cline{2-3} 
                                           & Gambling                    & \textit{\begin{tabular}[c]{@{}l@{}}"My bank card has been frozen due to internet \\ gambling. What should I do?" (live stream) \\ \textit{"Is it illegal to take back money lost in card games?"} \\ (short video)\end{tabular}}                                      \\ \hline
\begin{tabular}[c]{@{}l@{}}\\Social Laws \\ \\ \textit{These laws address various social issues,}\\ \textit{including family law, environmental law,}\\ \textit{and social welfare.}~\cite{chineselaws}\\ \\ \end{tabular}                             & Family Violence             & \textit{\begin{tabular}[c]{@{}l@{}}\\"My husband beats me at home. Should I get a \\ divorce?" (live stream)\\"Reasons for the persistence of domestic violence, \\its Scope, and self-help measures" (short video)\\\\\end{tabular}}                                                          \\ \hline
\begin{tabular}[c]{@{}l@{}}\\Administrative Laws  \\ \\ \textit{These regulations define the powers and}\\ \textit{functions of government agencies and}\\ \textit{cover the relationship between the}\\ \textit{government and the people.}~\cite{chineselaws}\\ \\ \end{tabular}                       & Buying Prostitutes          & \textit{\begin{tabular}[c]{@{}l@{}}\\"I was administratively penalized for buying a \\ prostitute. How can I keep this from people in my \\ workplace?" (live stream)\\"Why prostitution can not be legalized?" \\(short video)\\\\\end{tabular}} \\ \hline
\end{tabular}}
\resizebox{\textwidth}{!}{\color{black}\begin{tabular}{l}
$^a$The Supreme People's Court in China outlines seven categories of the National Laws: Constitution and Constitution-related Laws, \\Civil and Commercial Laws, Administrative Laws, Economic Laws, Social Laws, Criminal Laws, and Law on Lawsuit and Non-lawsuit \\Procedures~\cite{chineselaws}. We categorized the observed legal topics in live-streaming and short videos based on the Chinese law system and\\ provided corresponding definitions.
\end{tabular}}
\label{tab: live streaming}
\end{table}

\end{document}